\documentclass[10pt]{article}

\usepackage{epsfig}
\usepackage{amssymb}
\usepackage{graphicx}
\usepackage{color}
\usepackage{subfigure}
\usepackage[T1]{fontenc}
\usepackage{jheppub} 

\begin{document}

\baselineskip 6mm
\renewcommand{\thefootnote}{\fnsymbol{footnote}}


\newcommand{\nc}{\newcommand}
\newcommand{\rnc}{\renewcommand}



\newcommand{\tcb}{\textcolor{blue}}
\newcommand{\tcr}{\textcolor{red}}
\newcommand{\tcg}{\textcolor{green}}


\def\be{\begin{eqnarray}}
\def\ee{\end{eqnarray}}
\def\bea{\begin{eqnarray}}
\def\eea{\end{eqnarray}}
\def\nn{\nonumber\\}


\def\ct{\cite}
\def\la{\label}
\def\eq#1{(\ref{#1})}


\def\a{\alpha}
\def\b{\beta}
\def\g{\gamma}
\def\G{\Gamma}
\def\d{\delta}
\def\D{\Delta}
\def\e{\epsilon}
\def\et{\eta}
\def\ph{\phi}
\def\Ph{\Phi}
\def\ps{\psi}
\def\Ps{\Psi}
\def\k{\kappa}
\def\l{\lambda}
\def\L{\Lambda}
\def\m{\mu}
\def\n{\nu}
\def\th{\theta}
\def\Th{\Theta}
\def\r{\rho}
\def\s{\sigma}
\def\S{\Sigma}
\def\ta{\tau}
\def\o{\omega}
\def\O{\Omega}
\def\pr{\prime}


\def\half{\frac{1}{2}}

\def\goto{\rightarrow}

\def\na{\nabla}
\def\grad{\nabla}
\def\curl{\nabla\times}
\def\div{\nabla\cdot}
\def\pa{\partial}
\def\fr{\frac}

\def\bra{\left\langle}
\def\ket{\right\rangle}
\def\lb{\left[}
\def\lc{\left\{}
\def\ls{\left(}
\def\lp{\left.}
\def\rp{\right.}
\def\rb{\right]}
\def\rc{\right\}}
\def\rs{\right)}

\def\vac#1{\mid #1 \rangle}


\def\td#1{\tilde{#1}}
\def\check{ \maltese {\bf Check!}}


\def\Tr{{\rm Tr}\,}
\def\det{{\rm det}}
\def\text#1{{\rm #1}}


\def\bc#1{\nnindent {\bf $\bullet$ #1} \\ }
\def\ch {$<Check!>$ }
\def\ss {\vspace{1.5cm}}
\def\inf{\infty}

\begin{titlepage}

\hfill\parbox{5cm} { }

\vspace{25mm}

\begin{center}
{\Large \bf Quantum correlation in quark-gluon medium}

\vskip 1. cm
Chanyong Park$^a$\footnote{e-mail : cyong21@gist.ac.kr}
and Jung Hun Lee$^b$\footnote{e-mail : netic2@gmail.com}

\vskip 0.5cm

{\it  $^a\,$Department of Physics and Photon Science, Gwangju Institute of Science and Technology, Gwangju 61005, Korea}\\

{\it $^b\,$Department of Physics and Astronomy, Sejong University, Seoul 05006, Korea} \\

\end{center}

\thispagestyle{empty}

\vskip1cm


\centerline{\bf ABSTRACT} \vskip 4mm

We study thermodynamics and quantum correlations of the string cloud geometry whose field theory dual is the quark-gluon medium. We found the novel universality of the entanglement entropy first law in the high quark density limit. We also showed that a correlation function generally decreases as the entanglement entropy of the background medium increases due to the screening effect of the background. We study the UV and IR effects of the medium on phase transition behaviour observed in the holographic mutual information using both perturbative and numerical computations. Moreover, by numerical computation, we show that in the IR region the critical length obtained from the mutual information behaves similar to the correlation length of the two-point function.

\vspace{2cm}


\end{titlepage}

\renewcommand{\thefootnote}{\arabic{footnote}}
\setcounter{footnote}{0}



\section{Introduction}
In quantum many-body systems the entanglement entropy nicely characterizes the degrees of freedom for a pure state. This quantity provides one of the useful tools to figure out quantum phenomena both in high energy physics and in condensed matter physics. In the light of AdS/CFT \cite{Maldacena:1997re,Witten:1998qj}, the entanglement entropy can be interpreted as the geometric object called the minimal surface, which extends to the bulk AdS space \cite{Ryu:2006bv,Ryu:2006ef,Hubeny:2007xt}. This duality gives a specific relationship between geometrical objects and informational quantities of the quantum system, like entanglement entropy \cite{Takayanagi:2012kg,Li:2010dr,Fujita:2011fn,Ogawa:2011fw,Nozaki:2013vta,MIyaji:2015mia}, mutual information \cite{MolinaVilaplana:2011xt,Allais:2011ys,Molina-Vilaplana:2013xqa,Mukherjee:2014gia,Casini:2015woa,Ziogas:2015aja}, and complexity \cite{Brown:2015bva,Brown:2015lvg,Bernamonti:2019zyy,Bernamonti:2020bcf}. 

In this paper, we consider a deformed conformal field theory (CFT) with a nontrivial dual gravity theory called the generalized string cloud geometry, which consists of the black hole geometry and uniformly distributed open strings. Originally, the string cloud model has been proposed as a holographic dust model which was represented by the Einstein gravity coupled with a one-dimensional string instead of point particles \cite{Stachel:1980zr,Chakrabortty:2011sp,Chakrabortty:2016xcb,Park:2020jio}. Recently, this model has been paid attention to understanding various properties associated with quark-gluon plasma (QGP) \cite{Lee:2009bya,Park:2009nb,Jo:2009xr,Park:2011zp,Park:2017ray}, a strongly coupled thermal state of matter, produced in the collision experiment of heavy nuclei \cite{Chakrabortty:2011sp,Chakrabortty:2016xcb}. Our aim in this work is to investigate macroscopic and microscopic quantum correlations in the heavy quark-gluon medium.

For a system away from its critical point, the entanglement entropy can have nontrivial corrections associated with the subsystem size \cite{Calabrese:2004eu,Park:2015dia,Park:2015hcz,Kundu:2016dyk,Kim:2016jwu,Narayanan:2018ilr,Park:2018ebm,Park:2019pzo,Park:2020nvo}. In the entanglement entropy study, the subsystem size can be reinterpreted as the inverse of the energy scale. From the viewpoint of the renormalization group (RG) flow, unlike the momentum space RG usually utilized in a quantum field theory (QFT), the size of the subsystem, which is homologous to the minimal surface, is related to the real space RG flow used in the condensed matter physics. Generally, the RG flow crucially depends on the deformation we consider. In this work, we take into account two different deformations, the density of open strings and black hole mass.

We expect that the excitation increases the degrees of freedom of the underlying theory due to the screening effect between quark-gluon medium \cite{Kim:2007qk}. Due to this reason, the entanglement entropy with excitation increases along the RG flow in the UV region. We clarify this feature by utilizing holographic entanglement entropy of the deformed theory.

One of the most important properties of entanglement entropy is that it has information over correlation for the given system. A well-known example is the mutual information. This quantity consists of the linear combination of entanglement entropy. Another candidate which measures the correlation between two disjoint subsystems is the entanglement wedge cross section \cite{Takayanagi:2017knl}. This is defined as the minimal length of the cross section which divides the entanglement wedge into two parts. However, since for these quantities, there is a lack of direct comparison target in the dual field theory, we need to consider a more intuitive quantity that can be constructed in both the boundary and bulk theories. 

The entanglement entropy can be regarded as a quantum correlation because it quantifies measures of the correlation between two macroscopic quantum subsystems. In the semiclassical approximation, the two-point function can be defined by a Feynman path integral over the geodesic length of a massive particle propagating in the AdS space \cite{Susskind:1998dq,Balasubramanian:1999zv,Louko:2000tp}. Therefore, at least, for two-dimensional CFT the two-point function can be written in terms of the length of the minimal surface computed in the entanglement entropy calculation. More precisely, these two different measures are identically described by a one-dimensional geodesic curve anchored at the boundary two points. These boundary points coincide with the boundary of the entangling region for the entanglement entropy. Furthermore, it is also possible to identify two boundary points with the positions of two local operators for the two-point function. For this reason, by using the holographic renormalization technique \cite{Balasubramanian:1999jd,DeWolfe:2000xi,deBoer:2000cz,Bianchi:2001kw,Skenderis:2002wp} we can take into account a two-point correlation function of two local operators which corresponds to a microscopic quantum correlation.

Entanglement entropy depends on the regularization scheme due to the UV divergence structure. Mutual information can not only avoid this problem, which is UV finite, but it also provides a useful tool to capture a first order phase transition in the entanglement entropy \cite{Headrick:2010zt}. In particular, we survey the effect of the quark-gluon medium on phase transition observed in holographic mutual information using both perturbative and numerical computations.

 The rest of this paper is organized as follows. In section \ref{sec:2}, we review a holographic model associated with the string cloud geometry and its thermodynamic properties. In section \ref{sec:3}, applying the holographic formulation to deformed backgrounds, we study quantum corrections of the entanglement entropy due to the effects of excitation and compute two-point correlations in both UV and IR limits. In section \ref{sec:4}, we extend the discussion of section 3 to general d-dimensions. Further, we elaborate upon holographic mutual information in section \ref{sec:5}. Finally, we close this work with some concluding remarks in section \ref{sec:6}.


\section{Holographic dual of quark-gluon plasma\label{sec:2}}

In the holographic study, D-branes usually play an important role in making a nontrivial gravity theory and in describing its dual field theory. When $N_c$ $D3$-branes lie in a flat background  spacetime on top of each other, their gravitational backreaction allows a five-dimensional AdS space together with a five-dimensional sphere in the near horizon limit. The radius of the resulting AdS space is directly related to the number of $D3$-branes which on the dual field theory side corresponds to the rank of the gauge group $SU(N_c)$. The five-dimensional AdS in the Poincare patch further generalizes to an AdS black hole or black brane with an appropriate blackening factor
\be 
ds^2  = \fr{R^2}{z^2} \ls - f(z) dt^2 + \fr{dz^2}{f(z)} + \d_{ij} dx^i dx^j \rs ,
\ee
where the blackening factor is expressed in terms of a black hole mass $m$
\be
f(z) = 1 - m z^4 .
\ee
According to the AdS/CFT correspondence, the pure AdS geometry becomes the dual of a strongly interacting CFT at zero temperature, while the AdS black hole is the dual of a CFT at finite temperature. Using the holographic renormalization technique, the energy-momentum tensor of the dual CFT automatically vanishes at zero temperature. At finite temperature, on the other hand, the energy-momentum tensor is proportional to the black hole mass. In this case, the energy-momentum tensor is proportional to $N_c^2$ and traceless. This fact implies that the dual field theory of the black hole is associated with the excitation of massless adjoint matter like gluons. Therefore, we identify the black hole geometry with the gluon's excitation of the dual QFT.

Assuming that a probe $D3$-brane is located at the boundary, we can also take into account open strings connecting $D3$-branes at the center an the probe $D3$-brane. Since the end of an open string follows a fundamental representation, it on the probe brane plays a role of a fundamental matter like a quark with $N_c$ degrees of freedom. In this case, the fundamental matter has an infinitely large mass. Combining the above fundamental and adjoint matters allows us to investigate a quark-gluon plasma holographically. For example, various time-dependent quantities in expanding universes was studied in \cite{Park:2020jio,Park:2021wep}. In this section, we holographically investigate the entanglement entropy and two-point function of a local operator in the quark-gluon plasma.

To describe fundamental matter living on the dual field theory side, we first think of uniformly distributed open strings in a $(d+1)$-dimensional AdS space. The corresponding gravity theory can be described by the following action 
\be
S = \fr{1}{16 \pi G} \int d^{d+1} X \sqrt{-g} \ls {\cal R} - 2 \L \rs  + {\cal T}\sum_{i=1}^{N}  \int d^2 \xi \sqrt{-h} h^{\a\b} \pa_{\a} X^{\m} \pa_{\b} X^{\n} g_{\m\n}  ,
\ee
where $g_{\m\n}$ and $h_{\a\b}$ indicate a $(d+1)$-dimensional bulk metric and an induced metric on an open string, respectively. The first part including a negative cosmological constant $\L$ allows a $(d+1)$-dimensional AdS space, whereas the second part represents the action of open strings with a string tension, ${\cal T}$. This model has been known as the string cloud model \cite{Stachel:1980zr,Letelier:1979ej,Gibbons:2000hf,Herscovich:2010vr,Chakrabortty:2011sp,Chakrabortty:2016xcb}. Assuming that $N$ open strings are uniformly distributed, the above action leads to the following Einstein equation
\be			\la{eq:Einstein}
{\cal R}_{\m\n} - \half {\cal R} g_{\m\n} + \L g_{\m\n} =  T_{\m\n} ,
\ee
with 
\be
T_{\m\n} = - \fr{16 \pi G {\cal T} N}{V_{d-1}} \ \fr{\sqrt{-h}}{\sqrt{-g}}  \ h_{\a\b}  \ \pa^{\a} X_{\m} \pa^{\b} X_{\n} ,
\ee
where $V_{d-1}$ indicates a $(d-1)$-dimensional volume perpendicular to the open string's worldvolume. Taking a static gauge with $\xi^0=t$ and $\xi^1=z$, the stress tensor of open strings reduces to
\be
T_{\m\n}  = -  \fr{{16 \pi G \cal T} N}{V_{d-1}} \ \fr{\sqrt{-h}}{\sqrt{-g}}  \ h_{\a\b} \  \d^{\a}_{\m} \d^{\b}_{\n}.
\ee
Noting that $h_{\a\b} = g_{\a\b}$ for $\a, \b  = 0,1$ in the static gauge, $T_{\m\n} = 0$ for $\m, \n \ne 0,1$.  

The open string configuration we considered allows us to find the following analytic solution
\begin{equation}
ds^2 = \frac{R^2}{z^2} \left(-f(z) dt^2  + \frac{1}{f(z)} dz^2 + \d_{ij} dx^i dx^j\right),
\end{equation}
with a nontrivial metric factor 
\begin{equation}		\la{res:blackholefactor1}
f(z) = 1  -  \ta z^{d-1} - m z^{d} . 
\end{equation}
where $\ta$ is the uniformly distributed string cloud density and can be written as
\be
\ta = \fr{32 \pi G {\cal T}  N }{(d-1) R V_{d-1}} >0,
\ee
where we only consider the positive density $\tau>0$ for which the metric solution satisfies the weak and dominant energy conditions \cite{Chakrabortty:2011sp,Chakrabortty:2016xcb}. On the dual field theory side, one can identify $m$ and $\tau$ with the excitation energy of adjoint matter (gluons) and the density of fundamental matter (quarks), respectively.

Now, we investigate thermodynamics of the generalized string cloud geometry. To do so, it is more convenient to rewrite the black mass in terms of $\ta$ and the horizon, $z_h$,
\be
m = \fr{1}{z_h^{d}} - \fr{\ta}{z_h} .
\ee
Then, the Hawking temperature and Bekenstein-Hawking entropy read
\be
T_H &=& \fr{ d  - z_h^{d-1} \ta }{4 \pi z_h} , \nn
S_{BH} &=& \fr{ V_{d-1}}{ 4 G  z_h^{d-1} } .
\ee
Here, the positivity of temperature yields the upper bound of the black hole horizon
\be
0 \le z_h^{d-1} \le \fr{d}{\ta }  .
\ee
Intriguingly, this black hole solution allows two horizons. Similar to a charged black hole case, these two horizons become degenerate at zero temperature and lead to an extremal limit. In the extremal limit, the blackening factor has a double root like 
\be
f(z) = \ls z_h - z \rs^2 F(z) ,
\ee
where $F(z)$ is regular even at the horizon. The existence of the extremal limit becomes manifest due to the fact that the horizon at zero temperature is located at a finite distance and, at the same time, $f(z_h)$ and $f'(z_h)$ automatically vanish. In the extremal limit, the near horizon geometry reduces to $AdS_2 \times R^{d-1}$ similar to the charged black hole case.

To see more details, we consider a three-dimensional generalized string cloud geometry 
\be     \label{scgeometry}
ds^2  = \fr{R^2}{z^2} \ls - f(z) dt^2 + \fr{dz^2}{f(z)} +  dx^2\rs ,
\ee
with  
\begin{equation}	 
f(z) = 1  -  \ta z - m z^2 . 
\end{equation}
Denoting an event horizon as $z_h$, the black hole mass, as mentioned before, is rewritten as 
\be     \label{relation1}
m = \fr{1}{z_h^2} - \fr{\ta}{z_h} .
\ee
Then, the blackening factor 
\be
f(z) = 1  -  \ta z + \ls \ta z_h -1 \rs  \fr{z^2}{z_h^2}  ,
\ee 
allows two roots, the event horizon $z_h$ and  inner horizon $z_{in}$,
\be
z_{in} = \fr{z_h}{\ta z_h -1} . 
\ee

Recalling that the boundary and the center of the string cloud geometry are located at $z=0$ and $z=\infty$ respectively, the assumption that $z_h$ is the outer horizon implies $z_h \le z_{in}$. Since $\ta$ and $z_h$ must be positive, this constraint restricts the range of $\ta z_h$ to be $0 \le \ta z_h \le 2$. Now, we summarize the possible horizons relying on the parameter regions

\begin{itemize}

\item For $\ta z_h <1$ the inner horizon does not exist, so there is only one horizon at $z=z_h$. 

\item For $1 \le \ta z_h < 2$, there are two non-degenerate horizons and $z_h$ corresponds to the outer horizon. 

\item For $\ta z_h = 2$, the outer and inner horizons become degenerate, which corresponds to an extremal limit. 

\end{itemize}
Although the range of $\ta z_h > 2$ is not consistent with our previous assumption, there still exists a black hole solution. In this case, $z_h$ and $z_{in}$ exchange their role. In other words, $z_{in}$ plays a role of the outer horizon instead of $z_h$. Since the string cloud geometry is invariant under the exchange of the inner and outer horizons, a similar parameter dependence discussed before occurs. From now on, we focus on only the case of $0 \le \ta z_h \le 2$.

A black hole solution generally maps to a thermal system satisfying the thermodynamics law. The string cloud geometry we considered can be also reinterpreted as a thermal system. The regularity of the above metric at the outer horizon leads to the Hawking temperature
\be \label{2dtem}
T_H = \fr{2 - \ta z_h}{4 \pi z_h} .
\ee
Using this relation, we can also represent all thermodynamic quantities in terms of $\ta$ and $T_H$ via
\be \label{bhorizon}
z_h =  \fr{2}{4 \pi T_H + \ta} .
\ee

As mentioned before, the extremal limit with $\ta z_h = 2$ corresponds to the zero temperature limit. In the black hole geometry, the thermal entropy is described by the Bekenstein-Hawking entropy which is proportional to the area of the horizon
\be
S_{BH} = \fr{L R}{ 4 G  z_h } ,
\ee
where $L$ is an appropriately regularized volume in the $x$-direction. Since these two thermodynamic quantities must satisfy the first law of thermodynamics, the energy contained in the system is given by
\be			\la{res:inenergymrho}
E = \int T_H \ d S_{BH} =  \fr{L R}{ 16 \pi G  } \fr{ 1 - \ta z_h}{z_h^2} + E_0 ,
\ee
where $E_0$ is introduced as an integral constant which does not affect on the thermodynamics law. Since we are interested in a thermal system corresponding to a black hole, we need to identify the above energy with the thermal energy which must vanish at zero temperature. To interpret $E$ as a thermal energy, $E$ must vanish at zero temperature satisfying $z_h = 2/\ta$, which fixes the integral constant to be $E_0  = L R \ta^2 / (64 \pi G) $. Therefore, the thermal energy of this system finally becomes
\be         \la{Result:energy}
E =  \fr{L R}{ 64 \pi G  } \fr{ ( 2- \ta z_h )^2}{z_h^2}  = \fr{\pi L R}{ 4 G}  T_H^2 ,
\ee
which automatically satisfies the Stefan-Boltzmann law. For later convenience, we define the energy density $\r$ contained in the system
\be         \la{Result:energydensity}
\r = \fr{\pi c T_H^2}{6} ,
\ee
where $c=3R/2G$ is the central charge of two-dimensional CFT.

Applying various thermodynamic relation, the fundamental thermodynamic quantities derived above allows us to get more information characterizing this thermal system. First, the heat capacity of this system is given by
\be
C_V =  \fr{L R (2 - \ta z_h) }{8 G z_h} \ge 0 .
\ee
Since the heat capacity is always positive in the considered parameter regions, we can see that the thermal system derived from the string cloud geometry is thermodynamically stable. Second, this system has the following free energy and pressure
\be
F &=& - \fr{ L R}{ 64 \pi G  } \fr{ 4 - \ta^2 z_h^2}{z_h^2}  , \\ \nn
P &=&  \fr{ R}{ 64 \pi G  } \fr{ 4 - \ta^2 z_h^2}{z_h^2}  .
\ee
Now, let us discuss an equation of state parameter which is one of the quantities characterizing the quark-gluon medium. Recalling that the energy depends only on the temperature, the equation of state parameter reads
\be
\o \equiv \fr{L P}{ E} =  1 + \fr{\ta}{2 \pi T_H} .
\ee
For a two-dimensional QFT, $\o=1$  indicates that the theory is conformal. If we consider the case of $\ta=0$, the medium consists of only massless gluons and leads to a CFT, as expected. In the very high temperature limit with $T_H \to \infty$, the equation of state parameter again approaches $1$. This is because in the limit of $T_H \to \infty$ the effect of the quark mass is negligible. As a result, the limit of $T_H \to \infty$ describes a UV fixed point. At finite temperature, intriguingly, the above result shows that the equation of state parameter increases linearly with the quark density. In other words, the pressure becomes high in the dense medium and in the low temperature region.

\section{Quantum correlation in the quark-gluon medium \label{sec:3}}

In the previous section, we investigated thermodynamic properties of the boundary thermal system dual to the string cloud geometry. Based on these thermal properties, from now on, we study various quantum correlations in the quark-gluon medium living in a two-dimensional spacetime. First, we focus on the macroscopic correlation in the quark-gluon medium. To specify the macroscopic quantum correlation, we exploit many different quantities like the  entanglement entropy, relative entropy, and mutual information, etc. In this work, we focus on the entanglement entropy which describe the quantum correlation between the inside and outside of the entangling surface. Since it is not easy to calculate the entanglement entropy of an interacting system on the QFT side, we will use the celebrated holographic technique proposed by Ryu and Takayanagi (RT) \cite{Ryu:2006bv,Ryu:2006ef}. The RT formula claims that the entanglement entropy of a QFT can be determined by the area of minimal surface extending to the dual bulk geometry.
 
To calculate the entanglement entropy holographically, we first divide the boundary space into two parts, a subsystem and its complement. Assuming that the subsystem lies in
\be
- \fr{l}{2} \le x \le \fr{l}{2} ,
\ee
the corresponding entanglement entropy is described by a minimal surface extending to the dual geometry
\be
S_E = \fr{R}{4 G} \int_{-l/2}^{l/2} d x \ \fr{\sqrt{z'^2 + f }} {z \sqrt{f}} .
\ee
Now, we introduce a turning point. Denoting the turning point as $z_0$, it represents a maximum value which the minimal surface can reach. In other words, the minimal surface extends only to $0 \le z \le z_0$ in the dual geometry. Due to the invariance of the entanglement entropy under $x \to - x$, the turning point automatically satisfies $z(0) = z_0$ and $dz/dx=0$ at $x=0$.

In the above entanglement entropy, there exists a conserved quantity. Using this conserved quantity, we can represent the subsystem size $l$ and the entanglement entropy in terms of the turning point 
\be
l  &=& \int_0^{z_0} dz \fr{2 z}{\sqrt{f (z_0^2 - z^2)}}   ,   \la{Action:HEEl} \\
S_E &=& \fr{R}{2 G} \int_\e^{z_0} dz \fr{z_0}{z \sqrt{f (z_0^2 - z^2)}}  \la{Action:HEE} ,
\ee
where $\e$ is introduced as a UV cutoff which regularizes the UV divergence. Since the string cloud geometry has the nontrivial blackening factor, it is not easy to perform the above integrals exactly. Thus, from now on we take into account two limiting cases characterizing the UV and IR behaviors.

\subsection{In a UV region \label{sec3:1}}

First, we consider the case with a very small subsystem size which describes quantum correlation in the UV regime. On the dual string cloud geometry, this small subsystem size limit is realized by taking the limit satisfying $z_0 / z_h \ll 1$ and $\ta z_0 \ll 1$. After calculating the integral \eq{Action:HEEl} perturbatively in the small subsystem size limit, rewriting the turning point in terms of $l$ gives rise to
\be
z_0 = \fr{l}{2} \ls 1 - \fr{\pi }{16} \ta l -  \fr{2 \pi }{ c} \r l^2  + \cdots \rs ,
\ee
where the ellipsis indicates higher order corrections and the energy density $\r$ was defined in \eq{Result:energydensity}. Using this turning point, the resulting entanglement entropy becomes perturbatively
\be     \la{Result:2dimHEE}
S_E = \fr{c}{3} \ls  \log \fr{l}{\e}  + \fr{  \pi }{16}  \ta  l   +  \fr{\pi}{c} \r l^2  + \cdots \rs  .
\ee
This result shows that the UV entanglement entropy grows up as the subsystem size increases. At a given subsystem size, the entanglement entropy becomes large when the string density $\ta$ and the excitation energy $\r$ are high, as expected. Defining a critical subsystem size as
\be
l_c = \fr{c \ta}{16 \r},
\ee
we see that for $l \ll l_c$ the effect of the quark density becomes dominant and the entanglement entropy increases linearly with the subsystem size $l$. For $l \gg l_c$, on the other hand, the excitation energy becomes dominant and the entanglement entropy increases by $l^2$. 

In the above entanglement entropy \eq{Result:2dimHEE}, the first term, which is independent of $\ta$ and $\r$ corresponds to the entanglement entropy $S_0$ of the ground state. Therefore, the entanglement entropy caused by the quark density and the excitation energy is given by
\be
\D S_E = S_E - S_0 = \fr{c}{3} \ls  \fr{  \pi }{16}  \ta  l   +  \fr{\pi}{c} \r l^2  + \cdots \rs .
\ee
Recalling that, after subtracting the ground state energy, the internal energy of the subsystem $\D E$ is given by \eq{Result:energy}, we can define the entanglement temperature $T_E$ as
\be
\fr{1}{T_E} \equiv \fr{\D S_E}{\D E} = \fr{4  Gc}{3\pi R T_H^2} \ls \fr{\pi \ta}{16} + \fr{\pi \r l }{c} \rs  = \fr{c\, \pi}{48}  \ls   \fr{\ta}{\r }   + \fr{ 16}{c} l \rs .
\ee
It was well known that the entanglement temperature in the UV region is universally proportional to the inverse of the subsystem size \cite{Bhattacharya:2012mi,Casini:2015woa}. However, the entanglement entropy studied in the quark-gluon medium shows a different possibility from the known universality. For instance, if we consider the case of a small quark density ($\r l \gg \ta$), the entanglement temperature is inversely proportional to the subsystem size, as mentioned before. In the dense quark-gluon medium satisfying $\ta \gg \r l $, however, the entanglement temperature becomes independent of the subsystem size. Intriguingly, this feature looks universal. In other words, the same feature also appears in higher-dimensional cases, as will be show later.

Now, we consider a microscopic two-point correlator of a local operator in the quark-gluon medium. A two-point function is generally affected by the interaction with the background medium, so that we can exploit a two-point function as a measure probing the property of the background medium. Assuming a local scalar operator whose conformal dimension is denoted by $\D_O$, then the two-point function of this operator on the dual gravity side is described by a geodesic curve whose two ends are attached to two local scalar operators. Denoting the position of two local operators as $x_1$ and $x_2$, the two-point function is given by
\be
\bra O(x_1)  O(x_2) \ket = e^{- \D_O \, L(|x_1 - x_2|) /R } , \label{geolength}
\ee
where $L(|x_1 - x_2|$ is a geodesic length connecting two local operators. For a holographic three-dimensional geometry, the geodesic length and entanglement entropy are associated with each other 
\be
L(|x_1 - x_2|) =  4 G S_E (|x_1 - x_2|) ,
\ee
where the subsystem size $l$ is replaced by the distance of two local operators, $|x_1 - x_2|$. In the UV region, using the above relation, the two-point function of a scalar operator  becomes in the quark-gluon medium
\be \label{tpscg}
\bra O(x_1)  O(x_2) \ket \sim \fr{1}{|x_1 - x_2|^{2 \D_O}} \ls 1 -  \fr{  \pi \D_O }{8}  \ta  |x_1 - x_2|   -  \fr{2 \pi \D_O }{c} \r |x_1 - x_2|^2  \rs .
\ee
Comparing this result with the CFT one implies that the two-point function in the quark-gluon medium rapidly decreases with increasing the distance of two operators. This is due to the screening effect caused by the background matters. In a high density regime, the two-point function reduces by $\ta |x_1 - x_2|$, whereas it reduces by $ \r |x_1 - x_2|^2$ in the low quark density region.

\subsection{In an IR region \label{sec3:2}}

In the previous section, we discussed the macroscopic and microscopic UV quantum correlation in the quark-gluon medium. We showed that these two quantum correlations lead to two different behaviors depending on the quark density and the excitation energy. Now, we investigate IR quantum correlations in the same quark-gluon medium. The turning point $z_0$ approaches the horizon $z_h$ in the IR limit where the subsystem size diverges. The IR entanglement entropy, therefore, can be rewritten as
\be
S_E = \lim_{z_0 \to z_h} \ls \fr{R}{4 G z_0} l + \fr{R}{2 G z_0} \int_\e^{z_0} dz \fr{\sqrt{z_0^2 -z^2}}{z \sqrt{f}} \rs.
\ee
The second term contains a UV divergence. After an appropriate renormalization procedure, this second term becomes finite even in the IR limit. On the other hand, the first does not have a UV divergence but diverges in the IR limit. This implies that the first term gives rise to a leading contribution to the IR entanglement entropy. Moreover, the first term is exactly the same as the thermal entropy included in the subsystem. In the IR limit, the quantum entanglement entropy flows to a thermal entropy with small quantum corrections. The thermal entropy in the quark-gluon medium is represented as a function of the quark density and temperature 
\be
S_{th} = \fr{ (4 \pi T_H + \ta) R}{8 G} l   .
\ee 
From the entanglement entropy point of view, the temperature is well defined only in the IR limit. As a result, the microscopic two-point function in the IR regime reduces to
\be
\bra O(x_1)  O(x_2) \ket \sim e^{- \D_O (4 \pi T_H + \ta) |x_1 - x_2| / 2 }  ,
\ee
where the correlation length should be identified as
\be \label{ircolength}
\xi_c\sim \frac{1}{4\pi T_H+\tau}.
\ee
Due to the screening effect of the quark-gluon medium, the two-point function is exponentially suppressed in the IR regime. This suppression becomes fast as the temperature and quark's density increase, as expected.


\section{Higher dimensional string cloud geometries \label{sec:4}}

The previous study on quantum correlations can extend to higher-dimensional cases. To check the universality of the UV entanglement entropy discussed before,
we investigate the UV entanglement entropy in a $d$-dimensional quark-gluon medium holographically. In the higher-dimensional case, unlike the previous two-dimensional case, we can take several different shapes of the entangling surface. In this work, we focus on a strip-shaped entangling region because the universal feature we are interested in is independent of the shape of the entangling surface. To do so, we parameterize a strip-shaped region as
\be
-\fr{l}{2} \le x = x_1 \le \fr{l}{2} \quad {\rm and} \quad  -\fr{L}{2} \le x_i \le \fr{L}{2} ,
\ee
where $i$ runs from $2$ to $d-2$ and $L$ indicates an appropriated regularized size of a total system. In a $(d+1)$-dimensional string cloud geometry, the holographic entanglement entropy is determined by  
\be
S_E = \fr{R^{d-1} L^{d-2}}{4 G} \int_{-l/2}^{l/2} dx \fr{\sqrt{f + z'^2}}{z^{d-1} \sqrt{f}} ,
\ee
where the blackening factor is given by
\be
f = 1 - \ta z^{d-1} -  \ls \fr{1}{z_h^{d }}  - \fr{\ta}{z_h} \rs  z^{d}.
\ee
A conserved quantity similar to the previous case determines the subsystem size and the entanglement entropy in terms of the turning point $z_0$
\be
l = \fr{2}{z_0^{d-1} } \int_0^{z_0} dz \fr{ z^{d-1}}{\sqrt{f} \sqrt{  1 - ( z/z_0 )^{2 (d-1)}  }} , \\
S_E = \fr{L^{d-2} R^{d-1}}{2 G} \int_\e^{z_0} dz \fr{1}{z^{d-1} \sqrt{f } \sqrt{1 - ( z/z_0 )^{2 (d-1)} }} .
\ee

\subsection{In the UV region}

In the UV region satisfying $z_0 \ll z_h$ and $z_0 \ll \ta^{1/(d-1)}$, the entanglement entropy can be reexpressed as a perturbative expansion. For $d=3$, the resulting entanglement entropy becomes
\be
S_E = 
\frac{ R^2}{4 G  } \fr{A_1}{\epsilon}
-\frac{ R^2 \Gamma \left(\frac{3}{4}\right)^4}{4 \pi  G } \fr{A_1}{l}
+\frac{\pi ^2  R^2  }{24 G \Gamma                   
      \left(\frac{3}{4}\right)^4}  V_2 \tau 
+ \frac{\pi ^2  R^2}{64 G \Gamma \left(\frac{3}{4}\right)^4 } \fr{V_2 l}{z_h^3} ,
\ee
where $A_1=2 L$ and $V_2 = l L$ are the area and volume of the two-dimensional strip-shaped entangling region, respectively. Thus, the third and fourth terms in the above entanglement entropy are proportional to the quark number and excitation energy contained in the subsystem, respectively. For $d=4$, the entanglement entropy in the UV region becomes
\be
S_E = 
\frac{R^3}{8 G }  \fr{A_2}{\epsilon^2} 
-\frac{\sqrt{3} R^3 \Gamma
   \left(\frac{2}{3}\right)^5 \Gamma \left(\frac{5}{6}\right)^2}{ 2^{14/3} \pi ^2 G }   \fr{A_2}{l^2}
+\frac{\sqrt{3} \pi ^2  R^3  \Gamma
   \left(\frac{4}{3}\right)}{ 2^{11/3} G \Gamma \left(\frac{2}{3}\right)^5}   V_3 \ta
+ \frac{81  R^3 \Gamma \left(\frac{4}{3}\right)^4}{80 G   \Gamma \left(\frac{2}{3}\right)^5 }  \fr{V_3 l}{z_h^4}  ,
\ee
where $A_2 = 2 L^2$ and $V_3=l L^2$ correspond to the area and volume of the three-dimensional strip-shaped entangling region. Here, the parts, which do not relies on $\ta$ and $\r$, correspond to the ground state entanglement entropy. The effect of excitation and quark density can be extracted by subtracting this ground state entanglement entropy. Intriguingly, the gluon's excitation and quark density, regardless of the dimension of the underlying theory, give rise to the similar contribution
\be
\D S_E \equiv S_E - S_0 = \fr{L^{d-2} R^{d-1}}{G} \ls g_1 (\ta) \, l +  g_2 (\ta,\r) \, l^2 + \cdots \fr{}{} \rs  ,
\ee
where $S_0$ indicates the ground state entanglement entropy and $g_1$ and $g_2$ are functions of $\ta$ and $\r = E /V_{(d-1)} \sim 1/z_h^d$. This fact leads to two distinct universal features in the quark-gluon medium. Since the energy is an extensive quantity, it is proportional to the subsystem's volume ($\sim l L^{d-2}$). For example, the energy contained in the subsystem for $d=3$ is in the high density limit ($\ta \to \infty$)
\be
E = \fr{\pi R^2 T_H^2 \sqrt{\ta}}{6 \sqrt{3} G} V_2 ,
\ee
while it in the low density limit becomes
\be
E = \fr{8 \pi^2 R^2 T_H^3}{27 G} V_2 .
\ee

The entanglement temperature in the low density limit ($\ta \to 0$) reduces to
\be
T_E = \fr{d E}{d S_E} \sim l^{-1} .
\ee
This is the well-known universality of the entanglement temperature which is originated from the relativistic excitation energy. We can also take into account another case which allows a different universality. Taking a high density limit of $\ta \to \infty$, we can ignore the effect of $\r$ and the quark density $\ta$ becomes dominant. Similar to the gluon's excitation case, we define the entanglement temperature which represents the change of the entanglement entropy when the quark density varies. Then, the entanglement temperature gives rise to another universality in the high quark density limit
\be
T_E = \fr{d E}{d S_E} \sim l^0 .
\ee
This implies that the entanglement temperature is independent of the subsystem size which is independent of the dimension $d$. As a consequence, the entanglement temperature in the UV regime can show two distinct universality relying on the quark density.

\subsection{Correlation in the IR regime}

In the IR region, the subsystem has a very large size. Its size for $d=3$ is determined as
\be
l = 2 \int_0^{z_0} dz \fr{z^2}{\sqrt{f} \sqrt{z_0^4 - z^4}} ,
\ee
and the corresponding entanglement entropy reads
\be
S_E = \fr{L R^2 z_0^2}{2 G}  \int_\e^{z_0} dz  \fr{1}{z^2 \sqrt{f}  \sqrt{z_0^4 - z^4} }  .
\ee
The entanglement entropy can be rewritten as the following form
\be
S_E = \fr{R^2 l L}{4 G z_0^2} + \fr{L R^2}{2 G z_0^2} \int_\e^{z_0} \fr{ \sqrt{z_0^4 - z^4}}{z^2 \sqrt{f} } .
\ee
In the IR region with $l \to \inf$, the turning point approaches the horizon, $z_0 \to z_h$. In this IR region, the first term proportional to the size of the subsystem diverges. The second term also has a divergence proportional to $1/\e$. This divergence corresponds to the UV divergence appearing in the previous UV entanglement entropy. Ignoring this UV divergence, the second term has no more divergence even in the case of $z_0 = z_h$. Therefore, the first term gives rise to the leading contribution to the IR entanglement entropy. Intriguingly, this leading contribution is exactly the same as the Bekenstein-Hawking entropy. This fact implies that the entanglement entropy evolves to the thermal entropy along the RG flow.

We can also investigate a microscopic two-point function in the IR regime. Regardless of the dimension $d$, the turning point determines the distance between two local operators
\be
|x_1 - x_2 | = 2 \int_0^{z_0} \fr{z}{\sqrt{f} \sqrt{z_0^2 - z^2}}  ,
\ee
which is the distance measured at the boundary. The length of the geodesic curve extending to the dual geometry results in
\be
D = 2 R z_0 \int_\e^{z_0}  \fr{1}{z \sqrt{f}  \sqrt{z_0^2 - z^2}}   .
\ee
Rewriting the geodesic length as the following form
\be
D = \fr{R}{z_0}  |x_1 - x_2 | + \fr{2R}{z_0} \int_\e^{z_0} \fr{\sqrt{z_0^2 - z^2} }{z \sqrt{f}}  ,
\ee
we see that, when we ignore the UV divergence, the first term gives rise to a leading contribution in the IR regime ($z_0 \to z_h$ and $|x_1 - x_2 | \to \inf$).
As a consequence, we finally obtain the following IR two-point function
\be
\bra O(x_1) O(x_2) \ket = e^{- \D_O D /R} \sim e^{-2 \D_O |x_1 - x_2 |/z_h} .
\ee
This result shows that, when the distance of two local operators increases, the IR two-point function suppresses exponentially due to the screening effect caused by the background quark-gluon medium.

\section{Mutual information in the string cloud geometries \label{sec:5}}
In this section, based on the holographic point of view, we survey another quantum information quantity related to entanglement, which is called mutual information.
Mutual information measures how much the subsystems are correlated with each other and is defined as  
\bea \label{mi}
I(A;B)=S(A)+S(B)-S(A\cup B),
\eea
Where $S(A\cup B)$ is the entanglement entropy for a union of two subsystems. Due to the subadditivity of the entanglement entropy, the mutual information is always positive. 
When the mutual information becomes zero, $I(A;B)=0$, holographically the minimal surface suddenly changes its bulk configuration from a disconnected shape to a connected one. This feature allows to capture a phase transition in which the critical length is determined by the ratio of the size of the boundary subregions to their separation. 

In order to study the effects of our holographic model on the mutual information, we calculate this quantity for two disjoint subregions where their sizes and separation length are given by $\ell_1, \ell_2$ and $h$ respectively. Depending the relative size between subsystems and separation, the mutual information becomes
\bea
I(A;B)= \left\{ \begin{array}{crr}
S(\ell_1)+S(\ell_2)-S(\ell_1+\ell_2+h)-S(h)  & \,\,\,\mbox{for}  &  h\ll \ell_1, \ell_2,
\\
0 & \mbox{for} & h\gg \ell_1,\ell_2.
\end{array}\right.
\eea
It is obvious that the critical length $h_c$, where $I(A;B)=0$, depends on the two parameters $\tau$ and $\rho$ characterizing the string cloud geometry. For small values of these parameters one can perturbatively find the critical length as
\bea \label{cl}
h_c=(\sqrt{2}-1)\ell-\frac{\pi(\sqrt{2}-1)\tau}{16\sqrt{2}}\ell^2+\frac{\pi(c\pi(13-8\sqrt{2})\tau^2-2048\rho)}{1024\sqrt{2}c}\ell^3+\cdots,
\eea
where we assume the two subsystems have the same size, $\ell_A=\ell_B=\ell$ for simplicity. Depending on the values of $\tau$ and $\rho$, the required critical length has a non trivial size compared to the one in AdS. To show that, we consider the separation variable as a function of $\tau$ and temperature. We observe that for the extremal case $T_{H}=0$, the critical length has the maximum value. It implies that the upper bound of $h_c$ where a phase transition occurs is determined by turning off the effect of $\tau$ in the quark-gluon medium. Therefore, by turning up the temperature (or increasing the density of energy) the critical length decreases. 
Notice that the critical length is always shorter than the one in pure AdS because the screening effect stemmed from the quark-gluon medium.


The above result \eqref{cl} is only valid for the UV region where the minimal surface is localized in the AdS boundary. To see the IR effect of the mutual information, since there is no analytic form of the solution, it is needed to compute \eqref{mi} numerically for a given background geometry. To achieve these numerics, we first rewrite the horizon in terms of the function of the temperature $T_H$ and the density of string $\tau$ using \eqref{bhorizon}.

\begin{figure}
\centering
\subfigure[$T_H=0$]{\includegraphics[width=0.45\columnwidth]{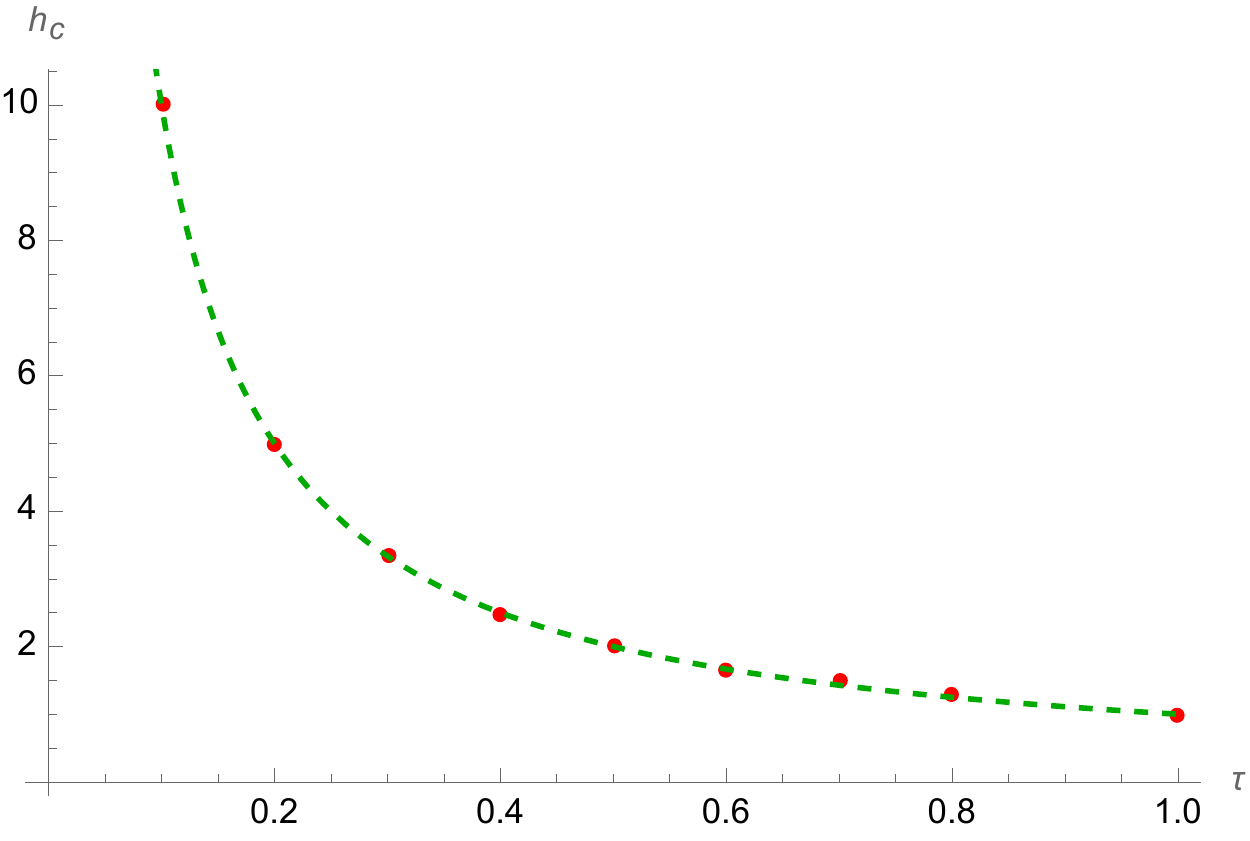}}
\hspace{0.5em}
\subfigure[$T_H=0.001$]{\includegraphics[width=0.45\columnwidth]{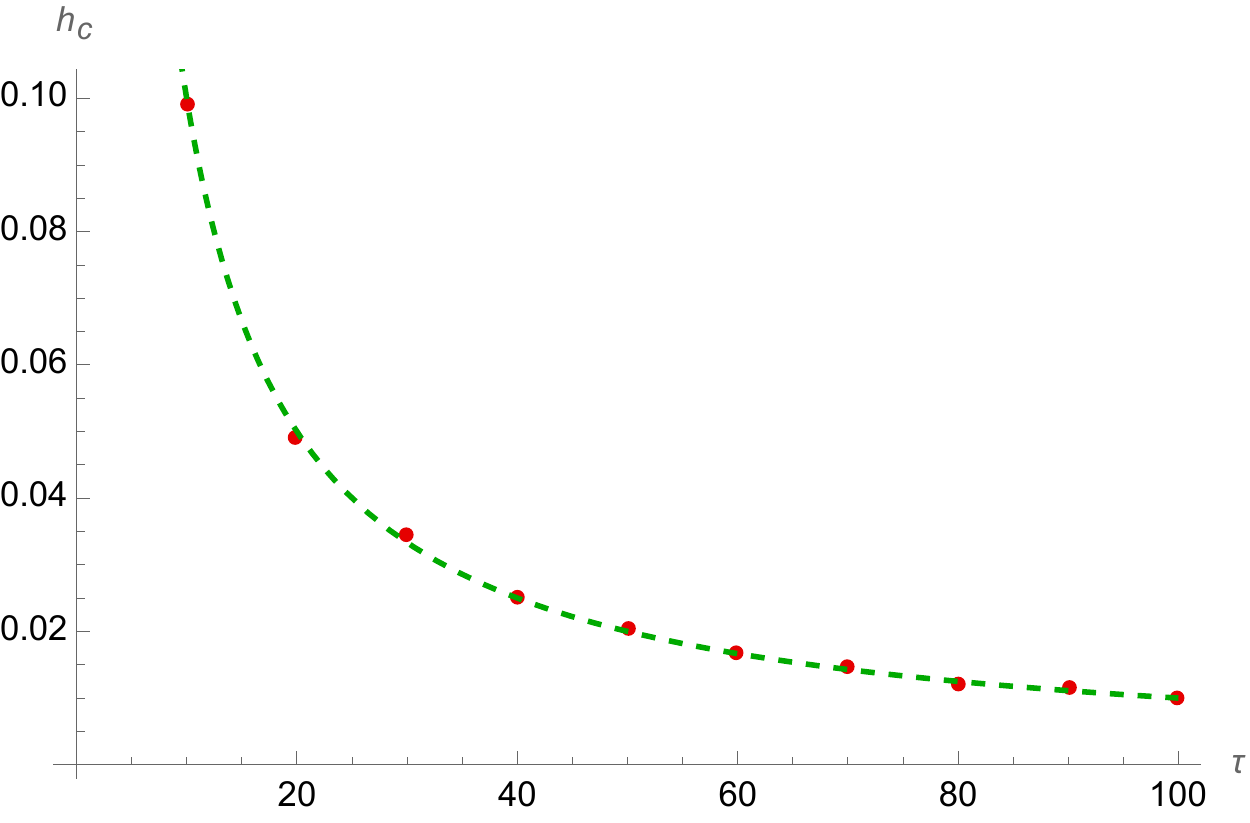}}
\caption{The plots of the critical length from $I(A;B)=0$ for the extremal black hole (a) and for the black hole with $T_H\ll \tau$ (b). The dashed green curve represents the microscopic correlation length $\xi_c$ in \eqref{ircolength} and the orange dots are numerical results.}
\label{colength}
\end{figure}

In section \ref{sec3:2} we studied the microscopic quantum correlation in an IR region by exponentiating the minimal length whose boundary points are assumed to be the positions of two local operators. For large values of the parameters, $T_H$ and $\tau$, a two-point function decays exponentially. The mutual information can also play a similar role in proving this kind of correlation. In this case, the mutual information measures a collective correlation between two subregions. From now, we call this the macroscopic quantum correlation. By numerical computation, we found that in the IR region the critical length $h_c$ behaves like the correlation length $\xi$ in \eqref{ircolength}, $h_c\sim \xi_c$, for both the extremal and large $\tau$ limits, see the Fig.\ref{colength}.




\section{Discussion \label{sec:6}}
Using the entanglement entropy, we holographically studied various quantum correlations in a quark-gluon medium. We considered a black hole geometry with a string cloud in order to represent heavy quarks and gluon excitations simultaneously. In the quark-gluon medium, quarks, gluons, and their interaction generate a nontrivial RG flow. Along this RG flow, quantum correlations usually vary and a UV theory flows to another IR theory. In this work, we investigated how a UV theory in a medium changes into a new IR theory by studying quantum correlations depending on the RG scale. At fixed points, a theory becomes scale invariant and has an infinite correlation length. However, a nontrivial RG flow generally breaks the scale symmetry given at fixed points. The scale symmetry breaking makes the correlation length finite. We looked into how the quark density and gluon's excitation energy modify the correlation length following the holographic propositions.

The entanglement entropy describes a quantum correlation between two macroscopic subsystems. We found that the entanglement entropy in the UV region always increases with increasing the quark density and gluon excitation. This fact indicates that the increase of the quark density and gluon excitation leads to increasing the degrees of freedom of the quark-gluon medium. As a result, the entanglement entropy grows up in a dense medium. We further took into account two local operators on this background medium and investigated their microscopic two-point correlation function. The entanglement entropy is governed by a minimal surface extending to the dual geometry, while a microscopic two-point function is described by a geodesic curve in the holographic setup. For a three-dimensional asymptotic AdS space, in particular, a minimal surface reduces to a geodesic curve. Therefore, the entanglement entropy of the background medium is closely related to a two-point function of local operators.  We studied a two-point function in the quark-gluon medium and found that a microscopic two-point function decrease when the quark density and gluon excitation grow up in the background medium. This result means that the higher the density and the higher the temperature, the stronger the screening effect of the background medium.

Another interesting thing in this work is the universality of the entanglement temperature in the medium. It was well known that the entanglement temperature reveals the universality in the UV region. For a relativistic theory, the entanglement temperature is inversely proportional to the subsystem size. In the quark-gluon medium, however, the entanglement temperature reveals two distinct universalities. In the low quark density limit where the gluon excitation becomes dominant, the entanglement temperature shows the expected universality, $T_E \sim 1/l$.  In the high quark density limit, on the other hand, we found that the entanglement temperature is independent of the subsystem size. Recalling that quarks we considered are heavy, we expect that the theory governing the dense medium deviates from a relativistic theory. Due to this reason, intriguingly, the entanglement temperature of a dense medium leads to a different universality, $T_E \sim l^0$.

We computed the holographic mutual information in the quantum system with the quark-gluon medium. A discontinuous change of the minimal surface can be interpreted as the first-order phase transition in the system. Utilizing both perturbative and numerical computations, it is shown that the critical length is diminished as the density and temperature increase which is the same as the results in Sec.\ref{sec3:1}. We also compare the microscopic IR correlation length with the critical length obtained from the mutual information in the IR region. By a numerical computation, we observed that in the IR region they are matched almost exactly.\\



{\bf Acknowledgement}

C. Park was supported by the National Research Foundation of Korea(NRF) grant funded by the Korea government(MSIT) (No. NRF-2019R1A2C1006639).  J. H. Lee was supported by the National Research Foundation of Korea(NRF) grant funded by the Korea government(MSIT) (No. NRF-2021R1C1C2008737).



	\bibliographystyle{apsrev4-1}

\bibliography{ref1}

\end{document}